# Real-Time State Modulation and Acquisition Circuit in Neuromorphic Memristive Systems

Shengbo Wang[1], Cong Li[1], Tongming Pu[1], Jian Zhang[1], Weihao Ma[1], Luigi Occhipinti[2], Arokia Nathan[3,4] and Shuo Gao[1]
[1]*School of Instrumentation and Optoelectronic Engineering, Beihang University, Beijing, China*
[2]*Department of Engineering, Univeristy of Cambridge, Cambridge, UK*
[3]*Darwin College, University of Cambridge, Cambridge, UK*
[4]*School of Information Science and Engineering, Shandong University, Qingdao, China*
shuo_gao@buaa.edu.cn

*Abstract*—Memristive neuromorphic systems are designed to emulate human perception and cognition, where the memristor states represent essential historical information to perform both low-level and high-level tasks. However, current systems face challenges with the separation of state modulation and acquisition, leading to undesired time delays that impact real-time performance. To overcome this issue, we introduce a dual-function circuit that concurrently modulates and acquires memristor state information. This is achieved through two key features: 1) a feedback operational amplifier (op-amp) based circuit that ensures precise voltage application on the memristor while converting the passing current into a voltage signal; 2) a division calculation circuit that acquires state information from the modulation voltage and the converted voltage, improving stability by leveraging the intrinsic threshold characteristics of memristors. This circuit has been evaluated in a memristor-based nociceptor and a memristor crossbar, demonstrating exceptional performance. For instance, it achieves mean absolute acquisition errors below 1 Ω during the modulation process in the nociceptor application. These results demonstrate that the proposed circuit can operate at different scales, holding the potential to enhance a wide range of neuromorphic applications.

*Keywords—neuromorphic system, memristor, bio-inspired system, artificial nociceptor, neural network*

## I. Introduction

Neuromorphic systems, with their dynamic nature similar to biological nervous systems, become strong candidates for developing bio-inspired applications [1]-[4]. Among the devices that make up neuromorphic systems, memristors stand out due to their high integration capability, low operation currents, and high similarity to biological synapses, particularly in bio-inspired applications like artificial tactile systems and artificial retinas [5]-[9]. Within these systems, a memristor's resistance (or conductance) not only stores previously set information but also adapts dynamically to external inputs, thereby adjusting the system's computational functions. Thus, modulation and acquisition of memristor states are crucial processes in managing the functions of memristive neuromorphic systems.

Currently, these two processes are separated in time, which introduces a significant delay that affects the real-time performance of neuromorphic systems [10]-[13]. For instance, during the setting of synaptic weights in memristive neural networks, write voltage pulses are followed by read pulses to modulate and then acquire memristor states, respectively. This sequential process increases the total operation time, thus hindering the system's real-time performance.

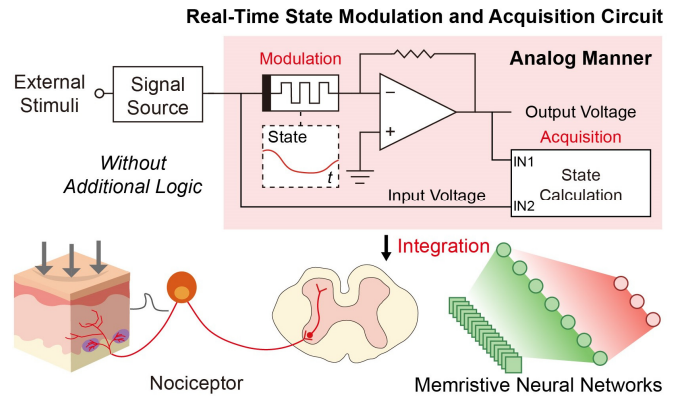

Fig. 1. The schematic diagram of the real-time state modulation and acquisition circuit and its integration with an artificial nociceptor and memristive neural networks.

In this work, we propose a dual-function circuit capable of concurrently modulating and acquiring the state information in memristive systems, as shown in Fig. 1. Utilizing the state maintaining characteristic of memristors without voltage stimuli exceeding the threshold, this circuit combines a feedback operational amplifier (op-amp) based circuit with a state calculation circuit that activates only when the modulation voltage exceeds the memristor's threshold voltage to achieve in-modulation state acquisition. To validate its effectiveness, we have integrated this circuit with an artificial nociceptor, achieving real-time status modulation and observation functionality with a mean absolute state acquisition error below 1 Ω. Furthermore, when integrated into a memristor crossbar, this circuit demonstrates a mean absolute acquisition error of 0.076 kΩ across 500 tests. The modulation voltage deviation remained at the mV level throughout the process. These results showcase that our method improves real-time capability while maintaining high accuracy, suggesting broad applicability across various neuromorphic systems.

## II. SDC Memristor and Modeling

The Self-Directed Channel (SDC) memristor, a metal ion-conducting device, is selected as the memristor unit for our dual-function circuit primarily due to its commercial availability and representative nature, as seen in its common switching mechanism. As shown in Fig. 2(a), the SDC memristor is a chalcogenide-based electrochemical metallization (ECM) device, a main type of switching mechanism in neuromorphic devices; its switching behaviors depends on the movement of $Ag^+$ ions into channels within the active layer, changing the device state. The electrical characteristics under tests are demonstrated in Fig. 2(b).

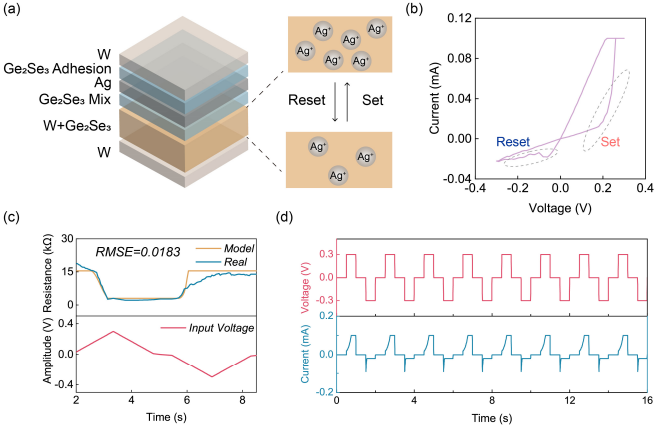

Fig. 2. The SDC memristor and its modeling. (a) The SDC memrsitor structure and its switching mechanism. (b) Hysteresis curve of the SDC memristor. (c) Comparsion of the real SDC memristor and its model. (d) Pulse response characteristics of the model.

Due to the need for observing the real state of memristors to further validate the state acquisition performance, the SDC is modeled in the SPICE simulator using the VTEAM model, a general model applicable to a broad range of memristor devices [14], [15]. The SDC memristor model defines the relationship between the external voltage stimuli and the state variable $w$ of the memristor. The derivative of the state variable is:

$$\frac{dw}{dt} = \begin{cases} k_{on}(\frac{v}{v_{on}}-1)^{\alpha_{on}} f_{on}(w), & 0 < v_{on} < v \\ 0, & v_{off} < v < v_{on} \\ k_{off}(\frac{v}{v_{off}}-1)^{\alpha_{off}} f_{off}(w), & v < v_{off} < 0 \end{cases} \quad (1)$$

where $v$ is the applied voltage stimuli, $v_{on}$ and $v_{off}$ are the threshold voltages, $k_{on}$, $k_{off}$, $\alpha_{on}$ and $\alpha_{off}$ are constants related to the resistance switching rate of the memristor, and $f_{on}(w)$ and $f_{off}(w)$ are the windows functions to preserve the memristor state variable $w$ within the physically realistic limits. Additionally, the relationship between the memristor resistance $R_M$ and state variable $w$ is exponential as follows:

$$R_M(w) = R_{on}(\frac{R_{off}}{R_{on}})^{\frac{w-w_{on}}{w_{off}-w_{on}}} \quad (2)$$

where $R_{on}$, $R_{off}$ and $w_{on}$, $w_{off}$ are the limit values of the memristance and the state variable $w$ in the ON and OFF state, i.e., the low resistance and high resistance state.

To fit the experimental data, the Broyden–Fletcher–Goldfarb–Shanno (BFGS) algorithm is utilized to optimize the model parameters with the Root Mean Square Error (RMSE) function defined as:

$$RMSE = \sqrt{\left(\frac{\sum((v_{model}-v_{real})^2)}{\sum(v_{real}^2)} + \frac{\sum((i_{model}-i_{real})^2)}{\sum(i_{real}^2)}\right)/N} \quad (3)$$

where $v_{model}$ and $i_{model}$ are the voltage and current calculated after optimizing the VTEAM fitting model, $v_{real}$ and $i_{real}$ are the real voltage stimuli and current responses, and $N$ is the sampling number of the $v_{real}$. The final fitting results achieve an RMSE of 0.0183, as shown in Fig. 2(c). Furthermore, Fig. 2(d) displays the voltage pulse response characteristics of the SDC memristor model. Detailed parameter values are available in Table I.

TABLE I. THE PARAMETERS OF SDC MEMRISTOR MODEL

| Parameter | Value | Parameter | Value |
|---|---|---|---|
| $R_{on}$ | 3.0 kΩ | $R_{off}$ | 15.5 kΩ |
| $\alpha_{on}$ | 1 | $\alpha_{off}$ | 1 |
| $k_{on}$ | 3.59 | $k_{off}$ | −4.83 |
| $v_{on}$ | 0.14 V | $v_{off}$ | −0.06 V |

## III. REAL-TIME MODULATION AND ACQUISITION CIRCUIT

The real-time state modulation and acquisition circuit consists of three primary blocks (Fig. 3): the feedback ammeter circuit, the voltage divider circuit, and the sample-and-hold circuit. During the memristor modulation process, the feedback ammeter circuit controls the voltage drop across the memristor terminals. The voltage divider circuit determines the memristor state based on the input and output from the feedback ammeter circuit. Simultaneously, the sample-and-hold circuit ensures stable and continuous operation of the divider circuit, even if the input voltage across the memristor changes crosses zero or becomes non-existent. By leveraging these blocks, the real-time memristor state modulation and acquisition can be archived. The detailed circuits are provided in the following sections.

### A. Feedback Ammeter Circuit

The structure of the feedback ammeter circuit is illustrated in Fig. 3. In this circuit, the positive terminal of the memristor is connected to the input voltage, while its negative terminal is connected to the inverting input of the operational amplifier. The non-inverting input of the op-amp is grounded, and its output is looped back to the inverting input. Owing to the existence of the virtual ground, the input voltage can be applied to the memristor without distortion, ensuring precise modulation of the memristor state. Moreover, the relationship between the input voltage $V_i$ and the output voltage $V_o$ can be expressed as:

$$V_o = -V_i \frac{R_f}{R_M} \quad (4)$$

where $R_f$ is the resistance of the feedback resistor and $R_M$ is the resistance of the memristor. When the feedback resistor value is given, the memristance $R_M$ can be derived from $V_i$ and $V_o$. This relationship is pivotal for the voltage divider circuit to determine the memristor state.

### B. Voltage Divider Circuit

The voltage divider circuit can be realized using a voltage multiplier within a feedback loop. In this configuration, $V_i$ and $V_o$ from the feedback ammeter circuit serve as the divisor and dividend, respectively. Therefore, whenever $V_i$ deviates from zero, the output of the voltage divider $V_q$, after changing the output voltage polarity using the op-amp, can be expressed as:

$$V_q = -\frac{V_o}{V_i} = \frac{R_f}{R_M} \quad (5)$$

However, during the memristor modulation process, there are instances where the input voltage is zero. It typically occurs when a sine wave transitions from its negative half-cycle to its positive half-cycle, or in the absence of any external voltage input. These scenarios can undermine the stability of the analog voltage divider, resulting in excessive output voltage and fluctuations. To ensure real-time and stable acquisition of the memristor state, the system integrates a sample-and-hold circuit with the voltage divider.

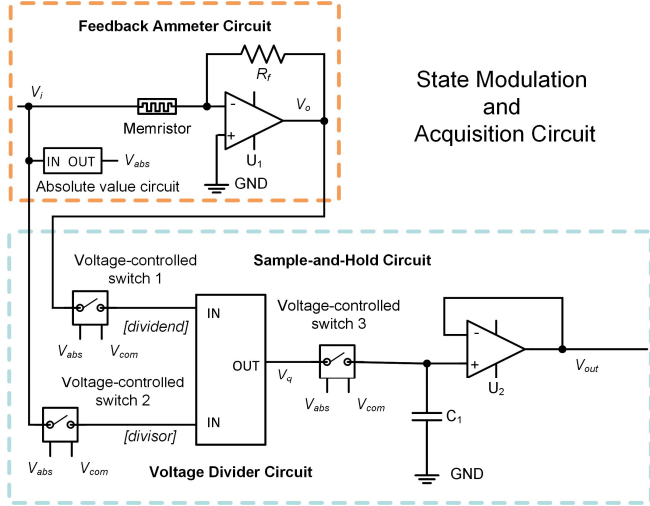

Fig. 3. The real-time state modulation and acquisition circuit.

*C. Sample-and-Hold Circuit*

As shown in Fig. 3, the sample-and-hold circuit primarily consists of voltage-controlled switches, a capacitor, and a voltage follower. The voltage-controlled switches are located at the input and output nodes of the voltage divider circuit and are regulated by the $V_{abs}$, the absolute value of input voltage $V_i$. Each switch establishes a connection from the feedback ammeter circuit to the divider circuit only when the difference between the $V_{abs}$ and the comparison voltage $V_{com}$ exceeds the opening threshold $V_{th}$ of the switch. Once the connection is established, the capacitor charges. When the analog switches are closed, the voltage of the capacitor mirrors $V_q$, representing the memristor state information. When the analog switches are turned off, the capacitor discharges extremely slowly owing to the high input impedance of the voltage follower. This means that the output of voltage follower $V_{out}$ will follow the $V_q$ when the switch is closed and maintain the previous voltage when the switch is open.

By integrating this sample-and-hold feature with the inherent threshold characteristics of the memristor, the voltage switches can be open when the input voltage lies between the positive and negative threshold voltages. This is because the state information of the memristor remains unchanged within this range. Consequently, challenges arising from a zero divisor voltage in the divider are negated, ensuring stable memristor state detection. For stable and accurate state information acquisition, the voltage $V_{sum}$ (the sum of $V_{com}$ and $V_{th}$) should be less than the smallest absolute value of the memristor's positive and negative threshold voltages, yet remain above zero. In this particular circuit configuration, $V_{com}$ and $V_{th}$ are fixed at 0 V and 0.05 V, respectively. This arrangement ensures the memristor's state information is updated when $V_i$ surpasses the threshold voltages of the memristor. If the absolute value of $V_i$ drops below $V_{sum}$, the state information of the memristor remains consistent, which is reflected in the output voltage $V_{out}$.

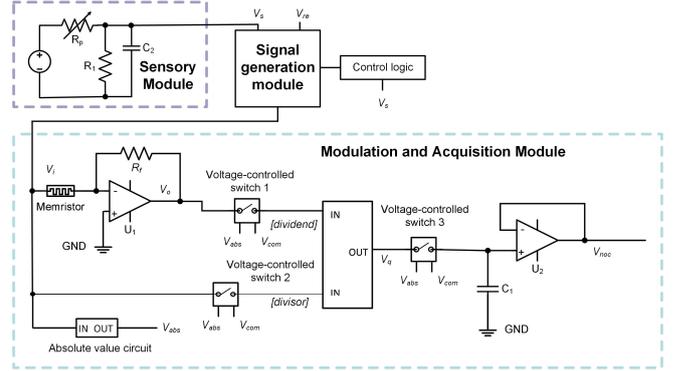

Fig. 4. The nociceptor with real-time status feedback functionality.

## IV. RESULTS AND DISCUSSION

*A. Nociceptor with Real-Time Observation Functionality*

Nociceptors, critical biological receptors for detecting harmful stimuli, have gained significant attention in bio-inspired systems, particularly in humanoid robots [16]-[18]. By integrating the state modulation and acquisition circuit, an artificial tactile nociceptor with real-time status observation functionality for humanoid robots has been demonstrated for the first time.

As depicted in Fig. 4, the nociceptor comprises three main blocks: the sensory module, the signal encoding module, and the modulation and acquisition module. In the sensory module, the resistance of piezoresistive film $R_p$ changes when force stimuli are perceived. Consequently, the voltage across the capacitor $V_s$ varies, representing the current intensity of the external stimuli. The signal encoding module, utilizing op-amps and voltage-controlled switches, generates modulation schemes based on the current stimuli intensity. When the intensity $V_s$ exceeds the sensory threshold of the nociceptor $V_{sth}$, the memristor in the modulation and acquisition module is modulated under the positive voltage $V_s$, and the increase in memristor conductance is analogous to the heightened sensitivity seen in biological nociceptors under harmful stimuli. Otherwise, the memristor is modulated by the negative recovery voltage $V_{re}$. During the processing of external stimuli, the virtual ground ensures the accurate voltage drop across the memristor while the modulation and acquisition module generates the voltage $V_{noc}$, representing the real-time status of the nociceptor.

Utilizing the circuit, three fundamental features of a nociceptor, 'threshold', 'no adaptation', and 'relaxation', are demonstrated via simulation. The voltage $V_{noc}$ represents the real-time observed memristor state, i.e., the sensitivity of the nociceptor, and $V_{noc}$ is directly proportional to the conductance of the memristor. As the memristor transitions from the high resistance state to the low resistance state, $V_{noc}$ shifts from 0.129V to 0.667V (with $R_f$ set to 2.0 kΩ). When the current stimulus is below the sensory threshold $V_{sth}$, the input memristor modulation voltage $V_i$ is the sum of $V_s$ and $V_{re}$. If the $V_i$ is below the inherent threshold of SDC memristor, the memristor state remain unchanged initially as shown in Fig. 5(a) and (b), consistent with the threshold feature observed in biological nociceptors. When the current stimulus exceeds the sensory threshold and $V_i$ exceeds the inherent voltage threshold of the SDC memristor, the memristor conductance increases. Thus, the observed state voltage $V_{noc}$ increases, indicating the nociceptor's enhanced perception (i.e., no adaptation) to hazardous stimuli.

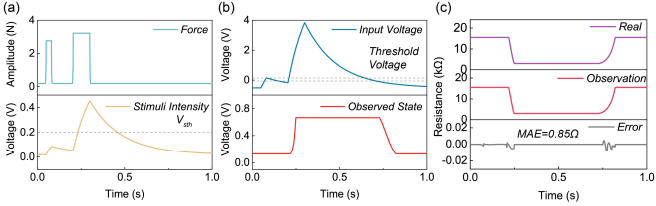

Fig. 5. The 'threshold' and 'no adaptation' feature of the nociceptor. (a) External force and stimuli intensity $V_s$. (b) Modualtion voltage $V_i$ for the SDC memristor and its observed state $V_{noc}$ based on our circuit. (c) State acquisition comparsion.

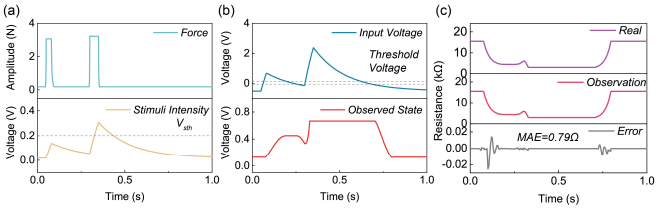

Fig. 6. Working performance in complex scenarios. (a) External force and stimuli intensity $V_s$. (b) Modualtion voltage $V_i$ for the SDC memristor and its observed state $V_{noc}$ based on our circuit. (c) State acquisition comparsion.

After the removal of external stimuli, the application of the recovery voltage returns the memristor to its original high-resistance state, and the observed state decreases, mirroring the relaxation feature seen in biological nociceptors. Comparing the resistance calculated from the observed state using Eq. 5 with the real resistance of the SDC memristor measured from the built-in voltage node in the simulation model, the mean absolute error (MAE) is 0.85 Ω, and the maximum absolute error is 8.32 Ω as shown in Fig. 5(c). Regarding modulation accuracy, the mean absolute voltage of the virtual ground of the op-amp connected to the memristor is only 1.55 μV.

In more complex scenarios, such as when a subsequent stimulus is applied before the nociceptor has fully relaxed, the nociceptor exhibits rapidly improved sensitivity, as shown in Fig. 6(a) and (b). The MAE and maximum absolute error in this case are 0.79 Ω and 25.11 Ω, respectively, as depicted in Fig. 6(c). Regarding modulation accuracy, the mean absolute voltage of the virtual ground is 1.34 μV.

*B. Memristive Neural Networks with Fast Setting Ability*

Integrating this circuit allows synaptic weights in memristive neural networks (MNN) to be set efficiently without the need for read pulses to determine the memristor states in the crossbar. As demonstrated in Fig. 7, a single circuit can manage the setting of a 4×4 layer in with selectors for the row and column of the setting synaptic unit.

In our validation tests, the write pulses consist of positive pulses (0.50 V, 50 ms) and negative pulses (-0.20 V, 50 ms) with a modulation interval set at 100 ms. We randomly select a synaptic unit and the polarity of the write pulses 500 times, and the modulation results observed through our circuit are displayed in Fig. 8(a). Throughout the entire setting process, the mean absolute voltage of the virtual ground is 0.12 mV.

When compared with the real memristor resistance, the MAE of the 500 tests is 0.076 kΩ (0.49% of the high-resistance state), and the maximum absolute error is 0.34 kΩ (Fig. 8(b)). The larger errors observed in these tests, compared to those in nociceptors, stem from the differences in circuits between single memristor and crossbar manipulation. These errors can be reduced by accounting for more circuit details in the crossbar, such as resistance in other paths within the crossbar. These findings demonstrate the potential of our circuit for application in large-scale neural networks to accelerate the setting of synaptic weights. To the best of our knowledge, this is the first instance of achieving concurrent memristor states modulation and acquisition, as shown in Table II.

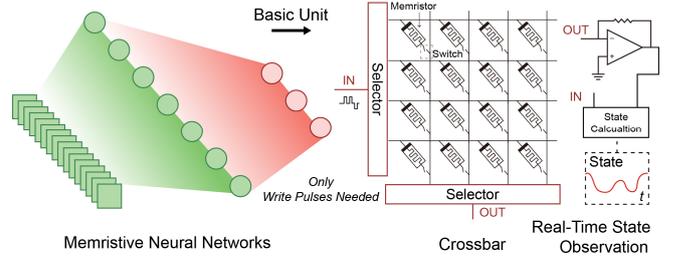

Fig. 7. Integration with memristive neural networks. The memristor crossbar in MNNs can be managed efficiently using the circuit.

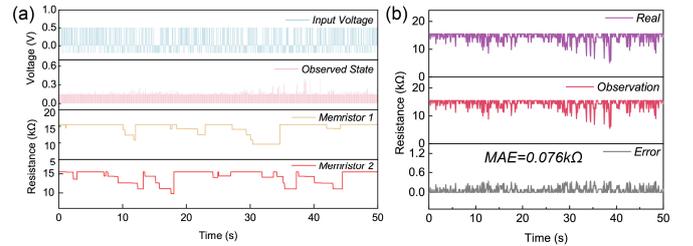

Fig. 8. Real-time observation results. (a) The generated input modualtion voltage and the observation result from the circuit. Real resistance change of memristor 1 at row 1 and column 1 and memristor 2 at row 1 and column 2, for example. (b) Comaprsion between the observation results and real corresponding memrsitor resistance.

TABLE II. COMPARISON WITH OTHER WORKS

| Nociceptor | Yoon et al. [19] | John et al. [20] | Zhu et al. [21] | Our Work |
|---|---|---|---|---|
| State Observation | × | × | × | √ |
| **MNN** | Yao et al. [12] | Reynolds et al. [22] | Wan et al. [23] | **Our Work** |
| Programming Method | Two Steps | Two Steps | Two Steps | Single Step |

## V. CONCLUSION

In this work, we have introduced a novel circuit that simultaneously modulates and acquires memristor state information in real-time by utilizing the inherent threshold characteristics of memristors. The circuit has been validated in artificial nociceptors and neural networks. Our design has the potential to streamline the implementation of biological mechanisms involving internal system state feedback and improve the real-time performance of advanced bio-inspired systems such as neuromorphic prosthetics and biomimetic robotics, advancing the development of the bio-inspired field [24]-[30]. To reproduce our results, the codes and materials used are available at https://github.com/RTCartist/Real-Time-Memristor-State-Modulation-and-Acquisition-Circuit.

ACKNOWLEDGMENT

S.G. acknowledges funding from National Key Research and Development Program of China (grant No. 2023YFB3208003), National Natural Science Foundation of China (grant No. 62171014), and Beihang University (grants No. KG161250 and ZG16S2103).